\documentclass[prl,twocolumn, preprintnumbers,amsmath,amssymb, superscriptaddress]{revtex4}
\usepackage{amssymb,amsfonts,amsmath}

\usepackage{mathrsfs}
\usepackage{graphicx}
\usepackage{dcolumn}
\usepackage{bm}
\usepackage{color}\usepackage{epstopdf}

\newcommand{\figa}{(a)}
\newcommand{\figb}{(b)}
\newcommand{\figc}{(c)}
\newcommand{\figd}{(d)}

\newcommand{\figf}{(f)}
\newcommand{\ffiga}{(a)}
\newcommand{\ffigb}{(b)}
\newcommand{\ffigc}{(c)}
\newcommand{\ffigd}{(d)}

\newcommand{\be}{\begin{equation}}
\newcommand{\ee}{\end{equation}}
\newcommand{\WI}{\bar W}

\newcommand{\Ps}{P(n)}
\newcommand{\Pm}{P_d(m)}
\newcommand{\Psm}{P(n, m)}
\newcommand{\Pscm}{P(n | m)}

\newcommand{\kb}{k_{\rm B}}
\newcommand{\mI}{\langle I \rangle}
\newcommand{\smI}{I}
\newcommand{\err}{\varepsilon_F}

\begin{document}

\title{ Experimental study of mutual information in a Maxwell Demon}

\author{J. V. Koski}
\affiliation{Low Temperature Laboratory (OVLL), Aalto University, POB 13500, FI-00076 AALTO, Finland}
\author{V. F. Maisi \footnote{Current address: \\ Solid State Physics Laboratory, ETH Z\"urich, 8093 \\ Z\"urich, Switzerland}}
\affiliation{Low Temperature Laboratory (OVLL), Aalto University, POB 13500, FI-00076 AALTO, Finland}
\affiliation{Centre for Metrology and Accreditation (MIKES), P.O. Box 9, 02151 Espoo, Finland}
\author{T. Sagawa}
\affiliation{Department of Basic Science, The University of Tokyo, Komaba 3-8-1, Meguro-ku, Tokyo 153-8902, Japan}
\author{J. P. Pekola}
\affiliation{Low Temperature Laboratory (OVLL), Aalto University, POB 13500, FI-00076 AALTO, Finland}

\begin{abstract}
We validate experimentally a fluctuation relation known as generalized Jarzynski equality (GJE) governing the work distribution in a feedback-controlled system. 
The feedback control is performed on a single electron box analogously to the original Szilard engine. 
In GJE, mutual information is treated on an equal footing with the thermodynamic work.
Our results clarify the role of the mutual information in thermodynamics of irreversible processes.
\end{abstract}


\maketitle


The second law of thermodynamics gives the inevitable upper bound of the available work that we can extract from fuels and heat baths.  
Information has  been recognized as another kind of thermodynamic fuel that can be converted into work with measurement and feedback control.
The relation between information and thermodynamics is a topic of long-standing interest in the field of statistical physics, dating back to the thought experiments of   ``Maxwell demon''~\cite{Maxwell,Szilard1929,MD}.
Relatively recent progress with universal nonequilibrium equalities applying to irreversible processes, known as fluctuation theorems~\cite{Bochkov1977, Jarzynski1997, Kurchan1998,Crooks1999,Jarzynski2000,Seifert2005}, has brought renewed attention to this problem.
In particular, the second law of thermodynamics and nonequilibrium equalities have been generalized to irreversible processes that involve information treatment, such as measurement, feedback control, or information erasure~\cite{Shizume1995,Piechocinska2000,Kawai2007,Esposito2011,Touchette2000,Sagawa2010,Sagawa2011,Sagawa2012,
Sagawa2013,Horowitz2010,Horowitz2011,Abreu2012,Horowitz2013,Ito2013}. A Maxwell demon is an object that measures the microscopic state of a system, and drives it to extract work or store energy with the aid of the measurement outcome.
A crucial element for the fidelity of this operation is mutual information $\mI$. It characterizes the correlation between the state of the measured thermodynamic system and the measurement outcome stored into the memory of the measurement device, and as such describes the efficiency of the measurement. 
Several recent experiments have illustrated the relation between information and thermodynamics~\cite{Toyabe2010,Berut2012,Berut2013,Koski2014}, however none have yet demonstrated the role of mutual information in irreversible feedback processes.

\begin{figure}[h!t]
\includegraphics[width=1\columnwidth]{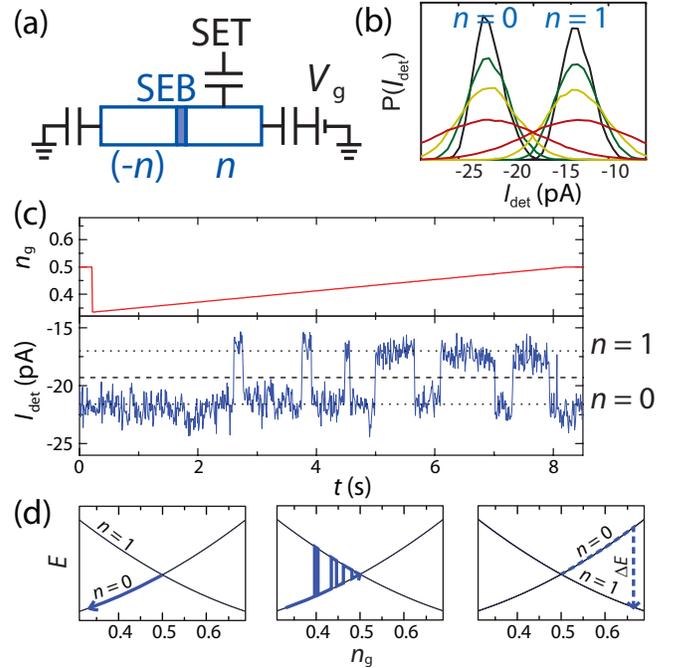}
\caption{Device and operation.\label{fig:device} \ffiga~Circuit diagram of the device. The single electron box (SEB), highlighted in blue, is the system under study. It is monitored by a single electron transistor (SET), whose current $I_{\rm det}$ depends on $n$, the number of excess electrons on the right island of the SEB. The SEB is controlled by gate voltage $V_g$. \ffigb~Single trace histogram of detector signal for states $n = 0$ (peaks to the left) and $n = 1$ (peaks to the right) with filter cut-off frequencies 50 (black), 100 (green), 300 (yellow) and 1000 Hz (red). \ffigc~The full trace of the feedback control. After the measurement and the rapid feedback, $n_g$ is brought quasistatically back to the original value. $I_{\rm det}$ shows the measured occupation in the SEB. \ffigd~Energy diagrams of the process. The rapid feedback (left) extracts work by lowering the energy. During the return back to degeneracy (middle), net work is extracted from the thermal excitations of $n$ entering the higher energy state. If the rapid feedback were performed incorrectly (right), excess work equal to $\Delta E$ would be applied to add energy to the system, and consequently the system immediately relaxes to the lower energy state. The return to degeneracy would again follow the behaviour of middle panel.}
\end{figure}

In this Letter, we study experimentally the mutual information in a feedback-controlled device. 
When the state of a generic thermodynamic system in state $n$  is measured with an outcome $m$, the stochastic mutual information~\cite{Sagawa2010,Sagawa2012,Sagawa2013,Abreu2012,Ito2013} is defined as
\be \label{eq:StochasticMutualInfromation} \smI (m,n) := \ln \Pscm - \ln \Ps, \ee
where $\Ps$ is the initial probability of the state being $n$, whereas $\Pscm$ is the probability that it is $n$ under the condition that the measurement outcome is $m$. Jarzynski equality (JE)~\cite{Jarzynski1997}, 
\be \label{eq:JE} \langle e^{-(W - \Delta F) / k_B T}\rangle = 1, \ee
has been generalized to systems with measurement and feedback control~\cite{Sagawa2010,Sagawa2011} to
\be \label{eq:GJarzynski} \langle e^{-(W - \Delta F) / \kb T - \smI}\rangle = 1, \ee
where $W$ is the applied work, $\Delta F$ is the change in free energy, $T$ is the temperature of the thermal reservoir, and $\kb$ is the Boltzmann constant. Equation \eqref{eq:GJarzynski} further reproduces the second law of thermodynamics as
\be \label{eq:SecondLaw} \langle W \rangle - \Delta F \geq -k_{\rm B} T \mI , \ee
where mutual information~\cite{Cover1991} $\mI$ is the expectation value of the stochastic mutual information. As $\mI$ is maximized in the ideal limit of the measurement correlating perfectly with the actual state, i.e. $P(n|m) = \delta_{mn}$, the magnitude of $\mI$ describes the efficiency of the measurement, providing the upper limit to how much work can be extracted from the system for the given information. 
We further define 
\be \label{eq:feedbackEfficiency}\eta_{\rm f} := \frac{-(\langle W\rangle - \Delta F)}{k_{\rm B} T \mI}\leq 1\ee
to describe the efficiency of the feedback control. If $\eta_{\rm f} = 1$, the feedback control is perfect and thermodynamically reversible, where all of the mutual information is extracted as work. The condition to achieve the reversible feedback has been discussed in Ref.~\cite{Horowitz2011}.

We perform the following experiment in a feedback-controlled two-state system. Our device is a single-electron box \cite{Averin1986, Buttiker1987, Lafarge1991} (SEB), illustrated in Fig. \ref{fig:device} \figa, which connects two metallic islands by a junction, permitting electron transport between the two by tunneling. The SEB is placed in a dilution cryostat, and the experiments are performed at $T = 100 \pm 3$ mK.
The two islands have a mutual capacitance $C_\Sigma$, such that tunneling electrons change the charge of this capacitor by elemetary charge $-e$ per electron. 
The charging energy of an SEB is 
\be \label{eq:chargingenergy} E(n, n_g) = E_C (n - n_g)^2,\ee
where $E_C = e^2 / 2 C_\Sigma$ is the energy required to charge the capacitor with a charge equal to a single electron, and $-en$ is the charge of the right island, induced by $n$ electrons that have tunneled from the left island. Our SEB has $E_C \approx 111~\mu$eV. Consequently, charge conservation requires that the charge of the right island is $en$. The electron tunneling is controlled by a nearby gate, accumulating a charge equal to $en_g = C_g V_g$ to the gate capacitor. 
The gate voltage $V_g$ is modulated to drive the SEB with $n$ being the stochastic state that changes by electron tunneling. 
The state $n$ naturally favours the energy minimum given by Eq. \eqref{eq:chargingenergy}, but can also change to a higher energy state due to thermal excitations.
The islands of the SEB are few $\mu$m long, providing a sufficiently small $C_\Sigma$ at sub-kelvin temperatures to achieve $E_C \gg k_B T$.
This diminishes the probability of higher energy states, practically limiting the SEB to be a two-state system with either $n = 0$ or $n = 1$ if we operate in the range $n_g = 0 ... 1$. While the SEB is modulated by the gate, a nearby single electron transistor (SET) monitors $n$. The measured trajectories of $n$ then determine the applied work $W = \int dt \frac{dn_g}{dt} \frac{\partial E}{\partial n_g}$. 

An SEB can be driven and monitored to test thermodynamic relations in a two-state system, and has already been used to verify various fluctuation relations \cite{Saira2012, Koski2013}. It can also be operated~\cite{Koski2014} as a Szilard engine~\cite{Szilard1929}, ideally extracting $\kb T \ln 2$ of work per feedback cycle. 
The steps of the operation follow the description introduced in Ref.~\cite{Horowitz2011}.
The initial energies of states $n = 0$ and $n = 1$ are equal by setting $n_g = 0.5$. In this degeneracy point, $n$ follows the distribution with equal probabilities $P(0) = P(1) = 1/2$.
The state $n$ is measured with the SET, providing an outcome $m$.
As feedback control, the gate is rapidly driven to $n_g = 0.5 \pm \Delta n_g$, where $\Delta n_g$ is a pre-determined parameter set to $\Delta n_g = 0.167$ for the present experiment, $+$ sign is used for $m = 1$, and $-$ sign for $m = 0$. This drive causes the state $m$ to have lower energy by $\Delta E = 2E_C \Delta n_g$ than the other state. Finally, $n_g$ is slowly brought back to degeneracy, extracting net work from concurrent thermal excitations of $n$. As this is a closed cycle with equal initial and final $n_g$, the free energy difference over the whole cycle is zero, $\Delta F = 0$, and we only need to consider $W$.

 \begin{figure*}[t]
\includegraphics[width=1\textwidth]{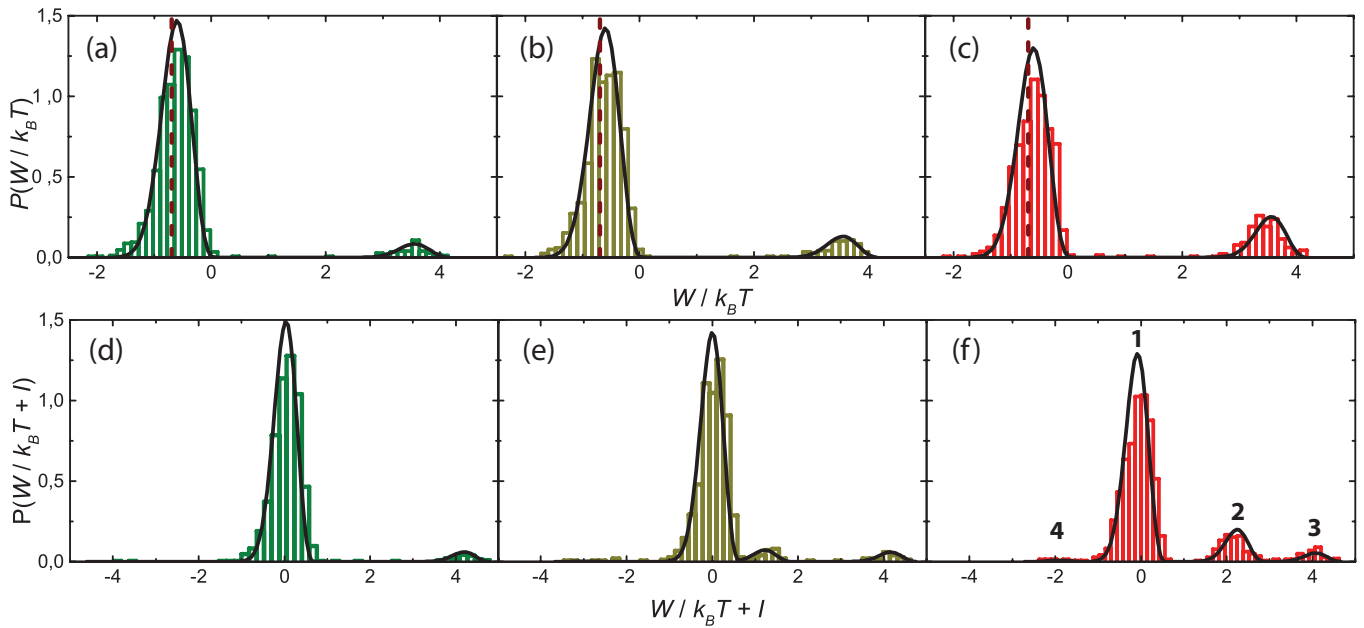}
\caption{Distributions of work with mutual information.  (\figa~- \figc) The measured $W / k_B T$ for different measurement error probabilities $\epsilon = 0.02, 0.05$ and $0.13$ with $N$ = 1177, 925 and $1025$, respectively. Black lines show the numerically obtained distributions. (\figd~- \figf) The distributions with the mutual information $I$ added to $W / k_B T$. The number in the panel~\figf~refer to the four peaks in Eq. \eqref{eq:PWI}.}
\label{fig:Distributions}
\end{figure*}

The mutual information is defined for our device as follows.
Let $\varepsilon$ be the error rate of the measurement, which is assumed to be equal for measuring $n = 0$ and $n = 1$; we obtain an incorrect outcome with probability $\Pscm = \varepsilon$ for $m \neq n$, while $\Pscm = 1 - \varepsilon$ for $m = n$.
If $\varepsilon = 0$, the measurement is error-free and $n=m$ holds.
The joint probability is given by $\Psm = P(m|n) \Ps$.
The probability of the detector measurement outcome $m$ is then written as $\Pm := \sum_n \Psm$, and the probability of $n$ conditioned on $m$ as $\Pscm = \Psm / \Pm$.  
At degeneracy, $P_d(0) = P_d(1) = 1/2$, implying that the Shannon information $S = -\ln \Pm$ written in the memory is $\ln 2$. 
Just after the measurement, $n$ and $m$ are correlated, which is quantified by the mutual information $\mI$.
By direct insertion to Eq. \eqref{eq:StochasticMutualInfromation}, we obtain stochastic mutual information $\smI(n, m) = \ln \left(2(1 - \epsilon)\right)$ for $m = n$, and $\smI(n, m) = \ln \left(2 \epsilon\right)$  for $m\neq n$.
The average of $\smI$ over all possible $n$ and $m$ produces the mutual information:
\be \label{eq:MutualInformation} \mI =    \sum_{nm} \Psm \smI(n, m) = \ln 2 + (1 - \varepsilon) \ln (1-\varepsilon) + \varepsilon \ln \varepsilon, \ee
which is the difference between the unconditional Shannon entropy $-\sum_{n}\Ps \ln \Ps$ and the conditional one $- \sum_{nm} \Pscm \ln \Pscm$.  Since the conditional entropy is not greater than the unconditional one, the mutual information is nonnegative.  It is also known that the mutual information is not greater than the Shannon information of the outcome; $\langle I \rangle \leq \ln 2$ holds in our setup, where the equality is achieved if the measurement is error-free (i.e., $\varepsilon = 0$).
The mutual information created by the measurement vanishes after the feedback control; if the feedback control is succesful, the mutual information is converted into useful work. We note that in order to complete the whole feedback cycle, the Shannon information written in the memory must be erased~\cite{Berut2012}.

To introduce the measurement error in the experiment, we change the cut-off frequency of the numerical low-pass filter applied to the detector signal for each measurement.
The signal for reading $n$ is subject to noise; increasing the cut-off frequency enhances noise in the filtered readout, as shown in Fig. \ref{fig:device}\figb.
This way, the probability $\varepsilon$ to measure the state $n$ incorrectly is different for each cut-off frequency.
In the analysis, for determining the trajectory of $n$ at the time of the measurement and during the foregoing feedback control, we filter the data with a low cut-off frequency (50 Hz), such that signal-to-noise ratio is high, and apply threshold detection as in Fig. \ref{fig:device}\figc. The tunneling rate at degeneracy, $\Gamma_0 = 2.7$ Hz, remains lower than the cut-off frequency of the filter and thus all the relevant transitions of $n$ are detected. The error probability $\varepsilon$ is extracted by counting the number of process cycles, where $n \neq m$. The feedback protocol is unchanged for each measurement.

\begin{figure*}[ht]
\includegraphics[width=\textwidth]{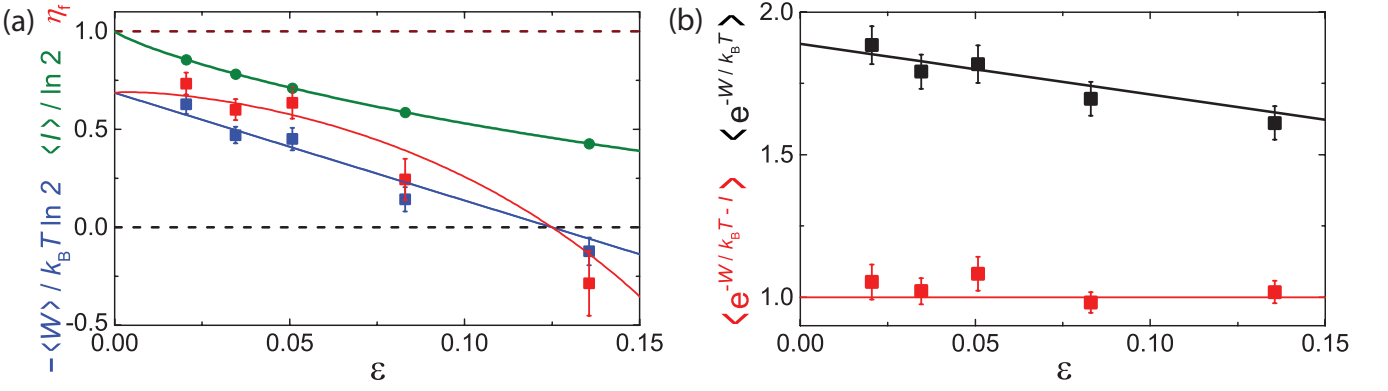}
\caption{Efficiency and fluctuation relations. \ffiga~Average extracted work (blue) and the feedback efficiency (red) as a function of measurement error. Experimental data are shown by symbols (in both panels), mutual information obtained from Eq. \eqref{eq:MutualInformation} is shown by green solid line, and numerical predictions for other quantities by solid lines (in both panels). The dashed brown line shows the fundamental maximum for $\eta_f$, $\langle I \rangle$, and $-\langle W\rangle$. The black dashed line shows the limit for $\eta_f$ and $-\langle W\rangle$ below which the process is dissipative. The results are obtained from $N$ = 1177, 1734, 925, 1060, and 1025 repetitions, in the order of increasing $\varepsilon$. \ffigb~Test of the generalized JE with mutual information. The measured exponential average of work with mutual information is close to unity, verifying the fluctuation relation within measurement accuracy. The error bars include the statistical error, as well as the uncertainity in the measured value of $E_C / k_B T$.}
\label{fig:Efficiencies}
\end{figure*}

Next we focus on feedback control. Ideally, $n_g$ is driven instantly after the measurement to obtain the energy difference $\Delta E$ between the states such that the initial state is at energy minimum. The work done in the fast drive is determined by the energy difference of Eq. \eqref{eq:chargingenergy}, $W_{0} = E(n, n_g \pm \Delta n_g) - E(n, n_g)$. After the fast response, the consequent slow return back to degeneracy practically starts from thermal equilibrium, as the rate of equilibriation is significantly faster than the rate of the drive. Let $P_0(W)$ be the probability distribution for applied work $W$ for an ideal fast feedback response followed by slow return to degeneracy. The slow drive satisifies JE \eqref{eq:JE}, and since the fast feedback response produces a fixed $W_{\rm 0}$, the distribution satisfies $\langle e^{-W / \kb T}\rangle_0 = 
e^{-(\Delta F_{\rm 0} + W_{\rm 0}) / \kb T} = 2 / (1 + e^{-\Delta E / \kb T})$, where $\Delta F_{\rm 0}$ is the free energy difference over the drive back to degeneracy, and $\langle...\rangle_0$ denotes averaging over $P_0(W)$. This condition allows us to determine the extracted work as $ -  \langle W\rangle_0 =  r\kb T \ln (2 / (1 + e^{-\Delta E / \kb T}))$, where $r$ is determined by the drive rate and has a value between $0$ and $1$. 

An error in the measurement produces an incorrect feedback response, making the distribution of work deviate from $P_0(W)$. However, measurement errors are not the only contribution to incorrect feedback responses. There is a finite delay of few ms between initiating the measurement, and triggering the feedback, which we call $\tau$. During this delay, the distribution of the states of the SEB evolves naturally, and the probability for the state to be different by the time the feedback takes place is $\delta = \frac{1}{2}(1- e^{-2\Gamma_0 \tau })$. In the presented experiment, $\tau \simeq 15$ ms. If either the measurement result is incorrect or the state has changed after the measurement, an additional $\Delta E$ work is paid as illustrated in Fig. \ref{fig:device}~\figd. These incidences thus follow a work distribution of  $P_0(W - \Delta E)$. On the other hand, if measurement is incorrect {\it and} the state changes after the measurement, the distribution again follows $P_0(W)$, and we obtain
\be\begin{split} \label{eq:PW} P(W) =~&((1 - \varepsilon)(1 - \delta) + \varepsilon \delta) P_0(W)\\ 
+~&(\varepsilon(1 - \delta) + (1 - \varepsilon)\delta) P_0(W - \Delta E), \end{split}\ee
matching the measured distributions shown  in Figs. \ref{fig:Distributions}~\figa~-~\figc.
Incidences with correct and incorrect measurement results have different values for $I$, and the resulting distribution modified by mutual information is
\be\begin{split} \label{eq:PWI} P(&\WI \equiv W + \kb T I) =\\&(1 - \varepsilon)(1 - \delta) P_0(\WI - \kb T \ln(2(1 - \varepsilon)))\\
+~&\varepsilon(1 - \delta) P_0(\WI - \Delta E - \kb T \ln(2 \varepsilon))\\ 
+~&(1 - \varepsilon) \delta P_0(\WI - \Delta E - \kb T \ln(2 (1 - \varepsilon)))\\ 
+~&\varepsilon \delta P_0(\WI - \kb T \ln(2 \varepsilon)),\end{split}\ee
which follows the generalized JE \eqref{eq:GJarzynski}.
The measured distributions shown in Figs. \ref{fig:Distributions}~\figd~-~\figf~match Eq. \eqref{eq:PWI}. In Fig. \ref{fig:Distributions}~\figf, the four peaks in the distribution are numbered in the order listed in Eq. \eqref{eq:PWI}.

The average extracted work given by the distribution of Eq. \eqref{eq:PW} is
\begin{equation}
- \langle W \rangle = r k_B T \ln \left(\frac{2}{1+e^{-\Delta E/ k_{\rm B}T}}\right) - \err \Delta E,
\label{eq:work_extraction}
\end{equation}
where $\err = \varepsilon (1 - \delta) + (1 - \varepsilon) \delta$ is the probability for incorrect feedback.
The extracted work is maximized by setting $\Delta n_g$ such that 
\be \label{eq:DeltaE} \Delta E / k_B T = \ln(r / \err - 1), \ee
with which, in the limits of $r \to 1$ and $\tau \to 0$, Eq. \eqref{eq:work_extraction} becomes an equality with Eq. \eqref{eq:SecondLaw}, as has been demonstrated in \cite{Horowitz2011}.
For any other $\Delta E$, $r$ or $\tau$, the extracted work is smaller in agreement with the second law of thermodynamics. In the ideal limit of $r \to 1$, $\varepsilon \to 0$, $\tau \to 0$, and correspondingly $\Delta E \to + \infty$, we obtain $- \langle W \rangle \to k_{\rm B}T \ln 2$ as is the case for the conventional Szilard engine. 

The generalized JE has also another form, which is given by~\cite{Sagawa2010}
\be \label{eq:GJarzynski2} \langle e^{-(W - \Delta F) / \kb T}\rangle = \gamma. \ee
Here $\gamma$ is a parameter that quantifies the efficiency of both the measurement and feedback.
While this equality has been verified in a colloidal system~\cite{Toyabe2010}, the present work is the first test of the generalized JE that connects thermodynamics and the mutual information, Eq. \eqref{eq:GJarzynski}.
The distribution of Eq. \eqref{eq:PW} produces
\be \label{eq:expW} \langle e^{-W / \kb T}\rangle = \frac{2}{1 + e^{-\Delta E / \kb T}} - 2\err \tanh\left(\frac{\Delta E}{2 \kb T}\right).\ee

Figure \ref{fig:Efficiencies} shows the measured expectation values discussed above as a function of measurement error. As one approaches the low-error regime, the incidences of incorrect feedback response become increasingly rare, and the average extracted work tends to approximately $0.7 \kb T \ln(2)$, as shown in Fig. \ref{fig:Efficiencies}~\figa. We find that the feedback efficiency $\eta_f$ given by Eq. \eqref{eq:feedbackEfficiency} remains almost constant for the lowest $\varepsilon$ and the extracted work primarily depends on measurement efficiency. For higher $\varepsilon$, the feedback protocol should be changed for a better $\eta_f$. The protocol could be optimized by correspondingly reducing the applied energy difference $\Delta E$ in accordance with Eq. \eqref{eq:DeltaE}.
Figure \ref{fig:Efficiencies}~\figb~shows the results for the test of the generalized JE. We see that Eq. \eqref{eq:GJarzynski} remains valid within measurement errors. 

In conclusion, our experiments illustrate the role of mutual information in the performance of a Maxwell demon. We show that our device follows generalized Jarzynski equality when under feedback control similar to that of a Szilard engine.  With a fixed feedback protocol, we show that the efficiency of the feedback changes with the measurement efficiency. 

This work has been supported in part by Academy of Finland (project no. 139172) and its LTQ (project no. 250280), the National Doctoral Programme in Nanoscience, NGS-NANO (V.F.M.), the European Union Seventh Framework Programme INFERNOS (FP7/2007-2013) under grant agreement no. 308850, and JSPS KAKENHI Grant Nos. 25800217 and 22340114 (T.S.). We thank O.-P. Saira for useful discussions. We acknowledge OMN, Micronova Nanofabrication Centre and the Cryohall of Aalto University for providing the processing facilities and technical support.

\end{document}